# Spot size dependent shock wave, plume and ion expansion dynamics of laser produced YBCO plasma

[1,2]S.C. Singh[#], [1]C.Fallon, [1]P.Yeates, C. McLoughlin[1] and [1]J.T. Costello

[1]National Centre for Plasma Science and Technology, Dublin City University, Dublin-9, Ireland

[#2]Present Address: Institute of Optics, University of Rochester, New York 14627, USA

**ABSTRACT**

The expansion dynamics of laser produced plasma plumes in gaseous atmospheres exhibit information on plasma-ambient gas interactions which result in plume splitting, shock formation, sharpening and confinement. We investigate laser spot size variation on shock wave, plume, and ion dynamics from laser produced $YBa_2Cu_3O_7$ (YBCO) plasmas using fast photography and Langmuir probe diagnosis. Changes in plume geometry are observed with varying focal spot size. At smaller spot sizes, lateral expansion of the plume is found to be larger, and plume expansion is spherical, while at larger spot sizes plume expansion is more cylindrical. Shock front formation time, relative intensity, spatial extent and total charge yield (TCY) are all dependent on laser spot size. Total charge yield (TCY) increases as the spot area increases, but decreases beyond a certain value. The width of the ion velocity distribution and the peak velocity decrease with increasing spot size, demonstrate that ions corresponding to larger spot sizes are somewhat more mono-energetic.

# I. INTRODUCTION

Laser produced plasmas (LPPs) are employed in basic research and industrial applications including pulse laser deposition (PLD) of thin films,[1] laser induced breakdown spectroscopy (LIBS) and laser ablation inductively coupled plasma mass spectroscopy (LA-ICPMS) for trace elemental detection,[2,3] preparation of nanoparticles in gaseous and liquid media,[4-8] laser surface nano/micro-structuring,[9] laser surface cleaning using shock wave peening[10] and micromachining.[11] An understanding of the numerous mechanisms which drive plume generation and expansion is central to these applications.[12-14] Detailed investigation of expansion dynamics provides information on the complex processes that occur during interactions between the plume and the ambient gas. This includes deceleration, attenuation, and thermalization of ablated species, shock wave formation and compositional evolution of ambient gas species in deposited films and nanostructures.[15-17]

The performance of $YBa_2Cu_3O_{7-\delta}$ (YBCO) thin films in high temperature superconducting (HTC) electronic applications depends on surface morphology, composition, and crystallinity which are governed by various interactions of the LPP plume with a background gas including scattering, attenuation, reaction,[18-20] plume geometry,[21] temperature, and the substrate-target distance.[22] These variables control the expansion dynamics of the ejected target material as it expands towards the substrate and can be configured to enhance the growth rate of films and nanostructures in PLD.[22-24] Electromagnetic confinement of the plume towards the target surface can enhance the rate of material deposition.[25] Physical confinement of the plume could also result in better collimation of ions to achieve higher deposition rate.[26] In PLD, the substrate surface is normal to the plume expansion vector, therefore a cylindrical plume configuration might be desirable compared to a hemispherical geometry.

The size, geometry and expansion dynamics of LPP plumes depend upon various target properties,[27] ambient atmosphere [18-20,23,28] and laser parameters.[22,29] Of these, laser irradiance is an important driver of the LPP characteristics and strongly affects expansion dynamics. It can be varied in two ways (i) changing the laser energy (spot size fixed), and (ii) changing laser spot size on the target surface (fixed laser energy). In case (i), with a fixed laser spot size, a relatively small fraction of increased laser energy is used for additional material vaporization from the target surface,[27] and the bulk of the energy is absorbed through laser-plasma interactions. However, case

(ii) requires variation of the spot size which changes the volume of ejected target material.[30] Lagrange *et al.* reported that at the same laser pulse energy, plume dynamics are quite different, illustrating that the above cases exhibit differing degrees of influence on LPP plume expansion dynamics.[31]

Singh and Narayan presented a three dimensional LPP expansion model to study influences of pulse energy density, target to substrate distance and laser spot size on spatial and compositional variations in deposited film.[32] Harilal *et al*. reported variations in the propagation dynamics of laser produced tin plasmas with laser spot size in argon at a pressure of 300 mTorr and observed sharpening of the plume front at larger spot sizes.[32] Mele *et al*. investigated the effects laser spot geometry on the spatial distribution of plume emission[33] while recently Li *et al.* studied the influence of laser spot size on the expansion dynamics of copper plasmas in an ambient atmosphere.[34]

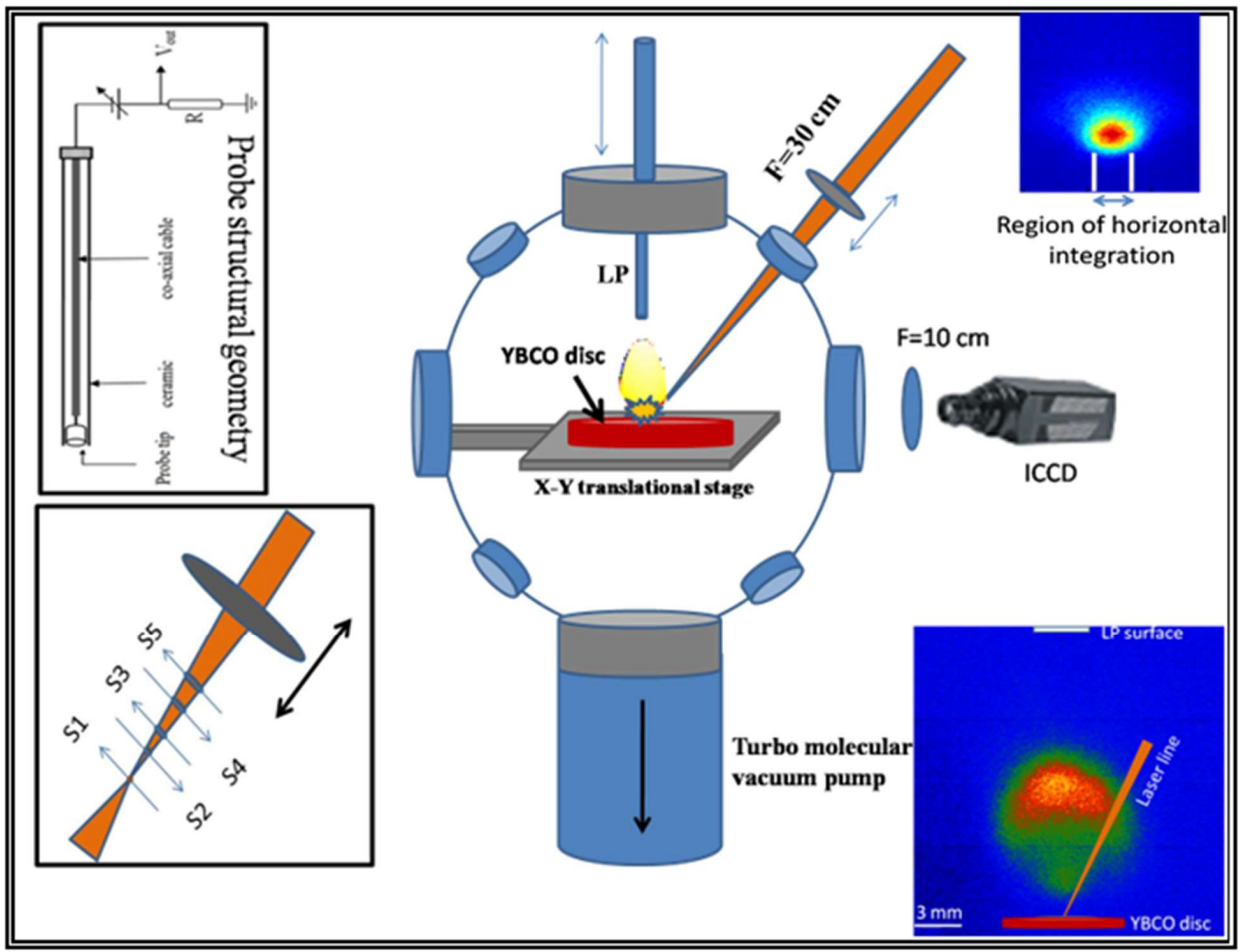

**FIG.1. Experimental arrangement for the plasma generation and optical imaging.** Left upper inset shows LP (Langmuir probe) structural geometry and the lower left inset illustrates focusing geometry to vary the focal spot size at the target surface. S1-S5 corresponding to the areas of spot sizes S1-S5. F: Focal length of convex lens, ICCD: Intensified Charge Coupled Device camera. Right upper inset shows plume region between two vertical lines was binned (parallel to the target surface) to illustrate the spatial variation of optical intensity along the direction of plume expansion.

In the present work, we explore the influence of laser spot size on the formation and expansion dynamics of laser produced YBCO plasma plumes, on shock wave formation and evolution during plasma plume-ambient gas interactions and on the total charge yield. Using high speed gated imaging and Langmuir probe ion detectors we diagnose plasma plumes produced with varying laser spot areas for a fixed laser pulse energy in an ambient air atmosphere (85 mBar). The measurements revealed that plume dynamics including size, shape, aspect ratio, sharpening, and splitting, time and intensity of the formation of shock front as well as ion current dynamics such as total charge yield and its full width at half maximum (FWHM), its peak height and velocity distribution exhibit a strong dependence on laser spot size.

The plume shape varies from hemispherical to cylindrical with increasing spot size. The plume lifetime ranged from 1600-2800 ns for on target fluence values of approximately 2-330 Jcm$^{-2}$ and the fluence also strongly determined the time and intensity of the shock front formation. The time index for shock front formation increased while shock front intensity decreased with increasing spot area. It is observed that total charge yield increases almost two-fold with an increase in spot area from 20-855×10$^{-4}$ cm$^2$, while it decreased for further increases.

## II. EXPERIMENTAL DETAILS

The experimental setup is depicted in Fig. 1. An YBa$_2$Cu$_3$O$_{7-\delta}$ (YBCO) target, 2.5 cm in diameter and 5 mm thick was mounted on a computerized X-Y motion control system at the centre of a spherical vacuum chamber, 40 cm in diameter, equipped with a turbo-molecular vacuum pump along with Penning and Pirani vacuum gauges. The target was translated in the horizontal plane to minimize the influence of cratering. Output pulses from a Spectron™, SL800 Nd:YAG laser (λ=1064 nm, 680 mJ, 16 ns) were directed at an angle of 45° downward from the vertical, into the chamber and focused *via* a plano-convex lens (30 cm focal length). The focusing lens was linearly translated on a linear rail to vary the average spot diameter from 260-3500 μm on the target surface (cf. lower inset of Fig. 1). Spots are ellipsoidal projections with semi-major and semi-minor axes which resulted in on target fluences from $F$=2 to 330 Jcm$^{-2}$. Table I summarizes the laser spot parameters. A digital delay generator (Quantum Composer™ 9500 Series) was used to trigger the laser and synchronize data acquisition for both the Langmuir probe, with traces captured *via* a fast oscilloscope and the intensified charge coupled device (ICCD, Andor Technology™;512×512 pixels; pixel size 24 μm). The gate time on the ICCD was set to 50 ns and images were acquired every 20 ns using a 10 cm focal length lens placed at 25 cm from the target center to image the expanding plasma plume onto the ICCD with 60 μm optical resolution. Wider gate width for the ICCD was chosen with the intention of capturing plume images for time indices up to 3000 ns after plasma ignition.

A planar Langmuir probe using a 15 mm tungsten tip and 3 mm in diameter was oriented at a fixed distance ($D$) of 2 cm from the target surface along the normal. The collecting surface was perpendicular to the direction of plume expansion (upper left inset of Fig. 1) and was biased at -78 V dc to repel energetic electrons. Data acquisition by the oscilloscope was triggered *via* a fast photodiode that intercepted scattered laser light from the target and was used to identify $T_0$ (TOF origin). The probe-target surface distance was varied and the corresponding Langmuir probe (LP) signals recorded by an oscilloscope (Tektronix TDS$^{TM}$ 3030 series).

**Table 1: Laser spot configuration S1-S5 parameters: major and minor axis of the laser spot, ellipsoidal spot area and resulting fluence ($F$) and power density ($I_P$) on target.**

| Laser Spot Configuration | Major axis (mm) | Minor axis (mm) | Spot area ($10^{-4}$ cm$^2$) | $F$ (Jcm$^{-2}$) | $I_P \times 10^9$ (Wcm$^{-2}$) |
|---|---|---|---|---|---|
| S1 | 0.31 | 0.21 | 0.21 | 330 | 20.0 |
| S2 | 0.66 | 0.47 | 0.98 | 70 | 4.40 |
| S3 | 1.96 | 1.39 | 8.60 | 7.9 | 0.50 |
| S4 | 2.7 | 1.9 | 16.0 | 4.2 | 0.26 |
| S5 | 4.1 | 2.9 | 37.0 | 1.9 | 0.12 |

## III. RESULTS AND DISCUSSION

### [A] Early Phase Imaging: Plume Generation and Expansion (<300 ns)

Gated imaging of an expanding LPP is a powerful diagnostic tool of the plumes interaction dynamics such as thermalization, splitting, and penetration of an ambient gas. This tool provides size and shape metrics of the plasma plume but also contributes to furthering the understanding of the plume propagation, plume edge hydrodynamics and reactive scattering at the plasma-gas interface. ICCD images of YBCO plasma plumes for different focal spot sizes, obtained at time delays of 20-300 ns after plasma breakdown are depicted in **Fig. 2.**

Initially, at a time delay of 20 ns, (ca. the end of the laser pulse) the electron density near the target surface is extremely high ($>10^{18}$ cm$^{-3}$).[43] Thus the plasma is highly collisional and the emission spectrum is continuum dominated.[43,44] The plume edge is determined using a $1/e^2$ criterion which defines the plasma interface as that coordinate were the recorded intensity drops to 13.5% of the recorded maximum[44]. This facilitates calculation of the expansion velocity ($V_{ESP}$) from ICCD images.

The LPP expands longitudinally and laterally with comparatively higher $V_{ESP}$ in the longitudinal direction. The plume configuration is approximately hemi-ellipsoidal with a bright expanding core and a relatively lower intensity annular periphery. For spot sizes S1 and S2, the LPP edge reaches 1.5 mm normal to the target surface in 20 ns, giving a longitudinal $V_{ESP}$ of 75 km/s. Increasing the spot size from S1 to S2 increases the radiant intensity and hence temperature of the LPP at the target surface with nearly equal plume sizes and peak radiant intensity. For spot sizes from S2 to S3, a decrease in plume $V_{ESP}$ (from 75 to 60 km/s), radiant intensity near the target surface and the size of the plume is observed. However, the peak radiant intensity remains almost unchanged. For S4 and S5, the peak radiant intensity, size of the plume and radiant intensity near the target surface are significantly reduced while $V_{ESP}$ decreases to 50 and 34 km/s respectively. Notably the position of peak radiant intensity for S4 is larger (1 mm) when compared to S5 (0.68 mm). At a time delay of 100 ns, the peak radiant intensity increases with increasing laser spot size from S1 to S2 and but then decreases monotonically. The spatial position of the peak radiant intensity is also maximized for the plume corresponding to S2 followed by S3, thus the size and temperature of the plume for S2 is highest at 100 ns.

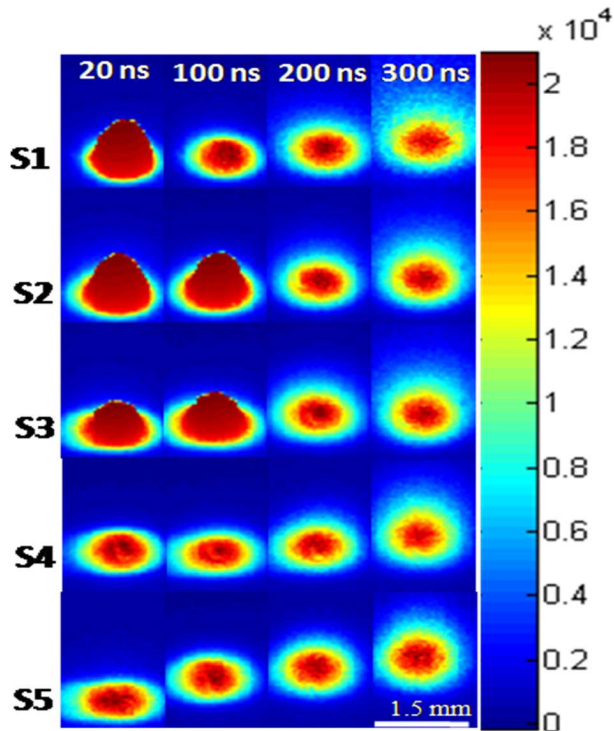

**FIG. 2.** Optical images of LPP plumes produced with different spot sizes from S1 to S5 and recorded with the ICCD at different time delays during the early phase ($\leq$ 300 ns) following the onset of plasma formation. Each image corresponding to a given spot size is obtained from different laser shots.

Spatial variation of the vertically binned (parallel to the target surface) emission intensity for the LPP plume images for all spot sizes from S1 to S5 at $t$= 200 and 300 ns delay are presented in figures 3(a) and 3(b) respectively. At $t$=200 ns, the peak radiant intensity of the plume monotonically increases with increasing focal spot size. The peak intensity was clearly a maximum for spot size S3, and decreased for S4 and S5. At $t$=300 ns the radiant intensity corresponding to S1 is a minimum while for S3 it is a maximum (Fig. 3(b)). The peak intensity dependence on spot size is identical for $t$=200 ns and 300 ns. Normalized peak radiant intensity *versus* time graphs for all laser spot sizes are presented in the inset of Fig. 3(a). The radiant intensity of the plume remains almost constant up to $t$ = 150 ns for laser spot sizes S2 and S3, while for S1 and S5 the radiant intensity decreases rapidly just after 20 ns. This suggests that plasma plumes produced with spot sizes S2 and S3 should exhibit higher plume temperatures for longer time histories compared to smaller (S1) and larger (S4 and S5) spot sizes.

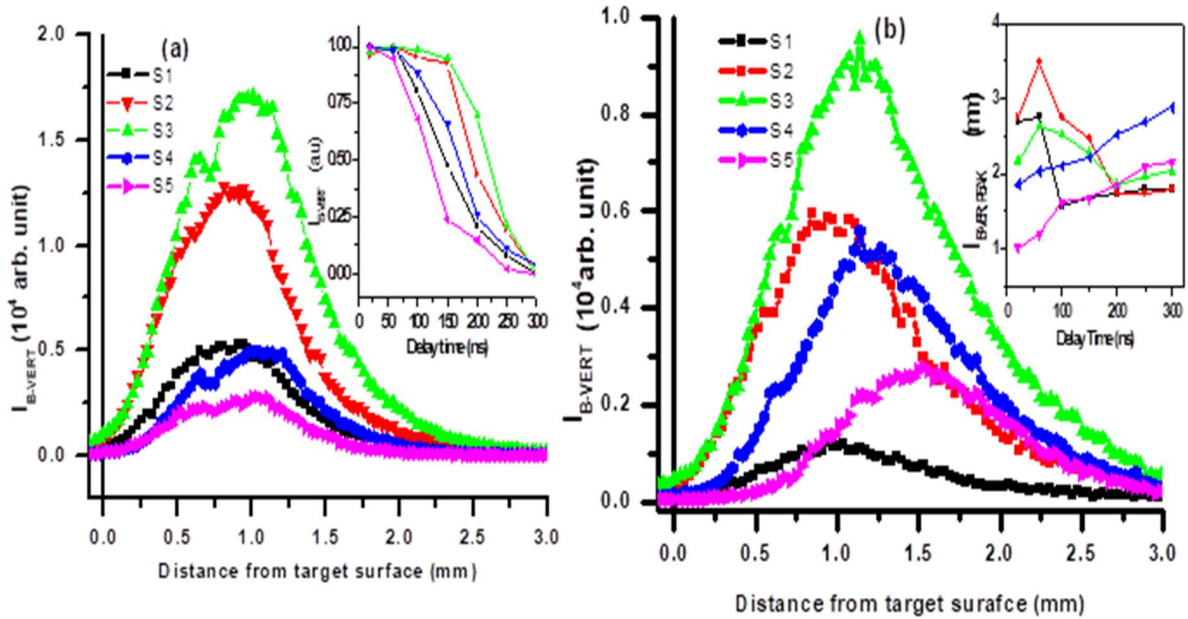

**FIG 3.** Influence of focal spot size on the spatial variation in the optical emission intensity of the plumes, shown in Fig. 2, along the plume expansion direction at (a) 200, and (b) 300 ns delays after the onset of plasma formation. The vertical line at zero corresponds to the target surface. Inset of (a): Temporal variation in the normalized scattered data of peak radiant intensity of the plume at different focal spot sizes from S1 to S5. Inset of (b): Temporal variation in the position of peak radiant intensity of LPP plumes produced with different focal spot sizes.

It is evident from the spatial position of the peak intensity *versus* time curves shown in the inset of Fig. 3(b) that for the laser spot sizes S1-S3, the positions of the peak intensity first increase with time, reach a maximum at 60 ns before decaying and this trend is spot size dependent. For S1, the position of peak radiant intensity of the plume decreases within 100 ns, while it decreases more slowly over a 200 ns period for S2 and S3. Thus in the first 60 ns, the intensity and plume temperatures attain their maximum values while, in the period 60-200 ns and for spot sizes S1 to S3, collisional interactions at the plasma-air interface reduces the plume's temperature. This process results in the peak intensity shifting from the plume edge to nearer the target surface leading to greater spatial compression of plume's radiant region. Since the internal plasma pressure greatly exceeds the ambient pressure, plume expansion continues until equilibrium is reached. For the delay times >100 ns in the case of S1 and >200 ns for S2 and S3, the position of peak radiant intensity increases slightly with time, which is evidence for a shift of the plasma-air interface away from target surface and hence greater plume expansion. For the spot sizes S4 and S5, the position of peak radiant intensity monotonically increases with time.

**[B] Intermediate Phase: Plume propagation, interaction, and splitting (400-1000 ns)**

In the intermediate phase, the interaction between the plume and ambient gas dominates resulting in the formation of a strong shockwave, reflection and scattering of plume species, ionization of ambient gas atoms/molecules by plasma species[15] and sharpening and splitting of plasma plume.[32] ICCD images of the YBCO plumes for all spot sizes for $t$ = 400 to 1000 ns are presented in Fig. 4. It is evident from the ICCD images that plumes produced with smaller spot sizes are more spherical, exhibit faster and stronger shock front formation but also a faster decay *i.e.* shorter emission lifetime. The spatial variations of the normalized emission intensities along the target normal, acquired from the ICCD images of Fig. 4, are illustrated in Fig.

5. For spot size S1 (fig.5(a)), a weak shock front is observed at 400 ns and the intensity of emission at the shock front increases with the increasing time delay. At later times $t_d \geq 600$ ns, the emission intensity from the shock region dominates, which is evident from the first row of ICCD images presented in Fig. 4 (S1). With increasing laser spot size, shock wave formation is first evident at longer delay times i.e. for spot size S1 shock formation occurs at approximately 400 ns, while this time rises to 600 ns for S2 and S3 and 900 ns for S4, while it is absent for S5At a given delay time, shock front intensity and its lateral expansion decrease while relative optical intensity of main plasma region increases with increasing spot size.

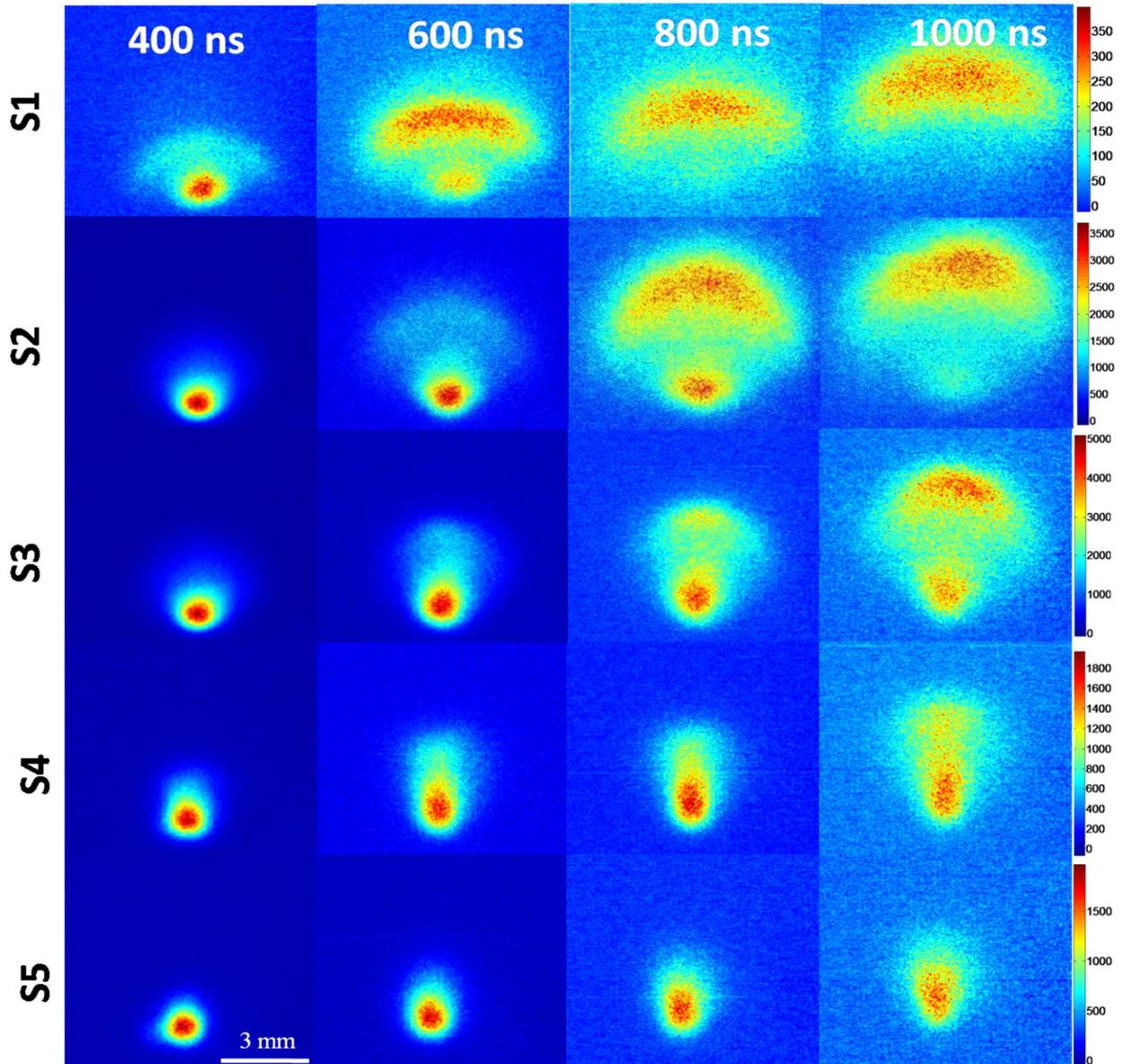

**FIG. 4.** Optical images of LPP plumes produced with different spot sizes from S1 to S5 and recorded with the ICCD at different times during the intermediate phase of plume expansion

*i.e.,* 400-1000 ns after the onset of plasma formation. Each image corresponding to given spot size is obtained from fresh surface.

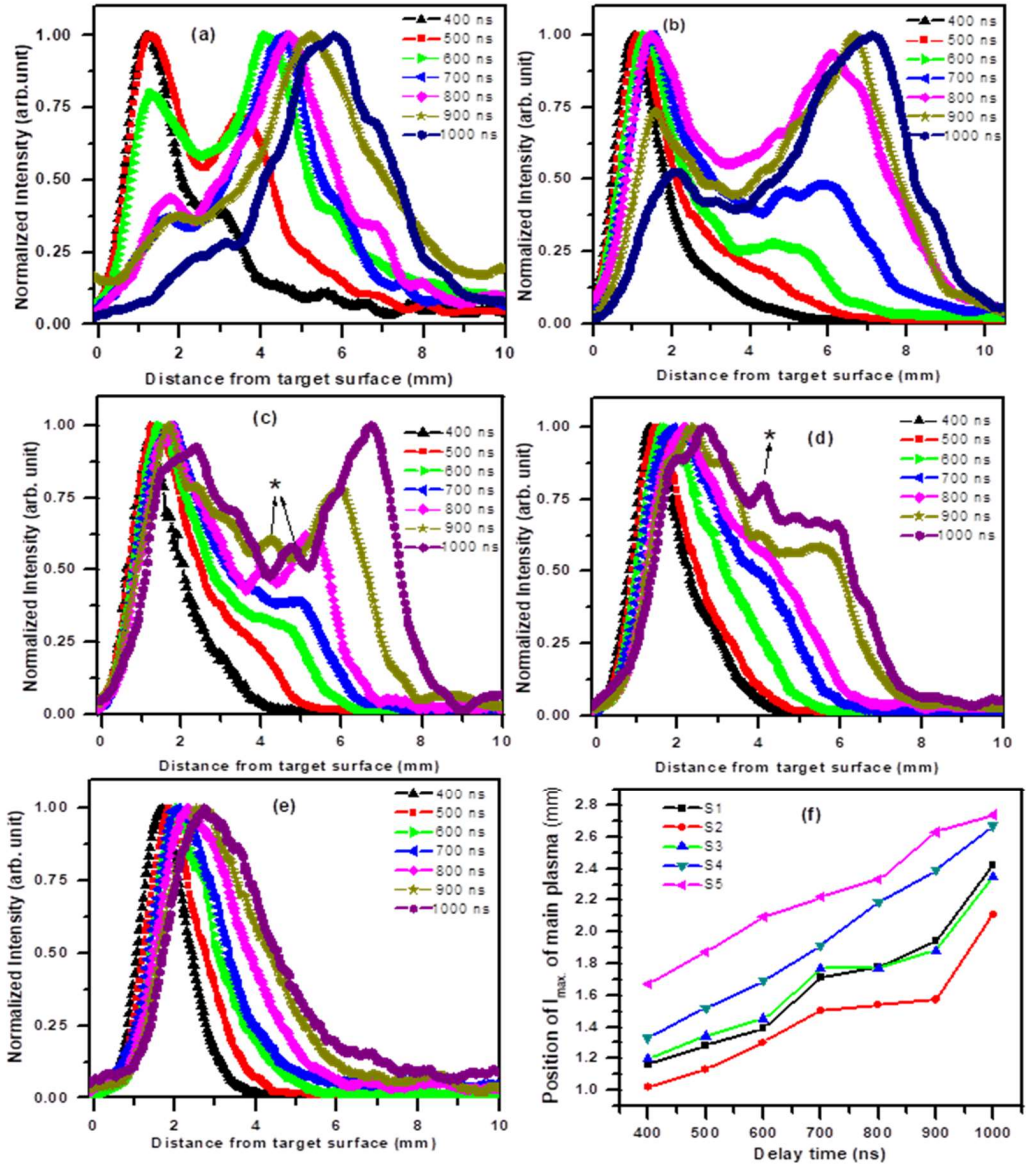

**FIG. 5.** Normalized optical intensity counts obtained from the ICCD images (Fig. 4) along the direction of plume expansion at various delays (400-1000 ns) after the onset of plasma formation for S1-S5 focal spot sizes. For better results, the intensity values from 25 pixels left to 25 pixels right on each side of the central axis of the plume are horizontally binned (inset C of Fig. 3). The resulting intensity profiles are normalized with respect to the maximum intensity for each focus condition. (f) Temporal variation in the peak position ($I_{max}$) of the main plasma corresponding to different focal spot sizes.

After the formation of the shock front, the time decay of the leading edge of the emission profile is sharper compared to the trailing edge (best observed for spot size S3), which demonstrates rapid deceleration of the plume front. A secondary emission intensity peak feature observed between the shock front and the rest of the plasma for delay times $t_d \geq 800$ ns and $t_d \geq 900$ ns corresponding to spot sizes S3 and S4 (marked with * in Figs. 5(c, d)) could be the consequence of scattering and reflection of the main plasma species backwards, from the shocked region, which in turn interact with the incoming species. At smaller spot sizes, the peak intensity of the shock region becomes larger than that of main plasma region at earlier times. For example: these times are 600, 900 and 1000 ns for the spot sizes S1-S3 respectively, and later at $t = 1400$ ns for spot size S4 (Fig. 5(d)), while there is no evidence of the presence of a shock front for spot size S5 (Fig. 5(e)).

The presence of a shock front also affects the intensity, emission lifetime and expansion dynamics of the main plasma species. The temporal variation in the position of the maximum intensity ($I_{max}$) of the main plasma, acquired from Figs. 5(a) to 5(e), corresponding to different spot sizes are shown in Fig. 5(f). It is evident from these plots that for the spot sizes S1-S3, where shock fronts are present, primary plasma constituents expand almost freely in the time window of 400-700 ns (weaker shock front), attain a plateau in the time range of 700-900 ns (strong shock front), followed by comparatively faster expansion at $t_d > 900$ ns. For larger spot sizes S4 and S5, where shock fronts are either weak or absent, the peak position of main plasma species varies linearly with time, reminiscent of free expansion in vacuum. Thus the shock front confines the plasma plume, which is evidenced from the smaller front position of main plasma species produced with smaller spot sizes (Fig. 5 (f)), where strong shock fronts are present.

### [C] Late Phase Imaging: Plume stopping and termination (> 1000 ns)

Internal pressure gradients within the plume drive expansion, normal to the target surface and the radiant emission from the plume decreases rapidly due to collisions with ambient air molecules at the plume boundary and radiative cooling *via* recombination processes. Under equilibrium conditions, the plume expansion rate decreases and the emission spectrum shifts from the soft X-ray and vacuum ultraviolet (VUV) to the visible region. Plume images at $t = 1200$, 1600 and 2000 ns for spot sizes S1-S5 are presented in Fig. 6. For smaller spot sizes S1 and S2, only shock fronts are visible during the period 1200-2000 ns. For S4 and S5, emission from the shock front in combination with the main plasma is present while for S5 only the main plasma is present. Spatial variation in the emission intensities along the primary expansion direction of each plume for *t*=1200 to 2000 ns are presented in Fig. 7. For plume images obtained at for S1, the emission profile is typical of a single emitting structure for *t*=1200 to 2000 ns. A weak two-body emission profile is however present at 1200 ns for S2 (Fig. 7(b)), but this is not present for later time delays. For a given ICCD gate width of 50 ns, the visibility of plume images was a maximum for S3 (Fig. 7(c)) up to 2400 ns, and a secondary trailing structure was observed nearer the target surface. The peak intensity of this slower plume region was ~50% of the primary peak near the plume's front for early delay times (1200-1400 ns). For $t$=1200 to 2000 ns plumes produced with spot size S4 exhibit a two zone emission structure comprising the main plasma species and a shock front. The emission intensity of the shock front increases while the main plasma intensity decreases with time. For spot size S5, plume images are visible up to 1500 ns, and the corresponding spatial distribution curves are single peaked with no discernible shock wave.

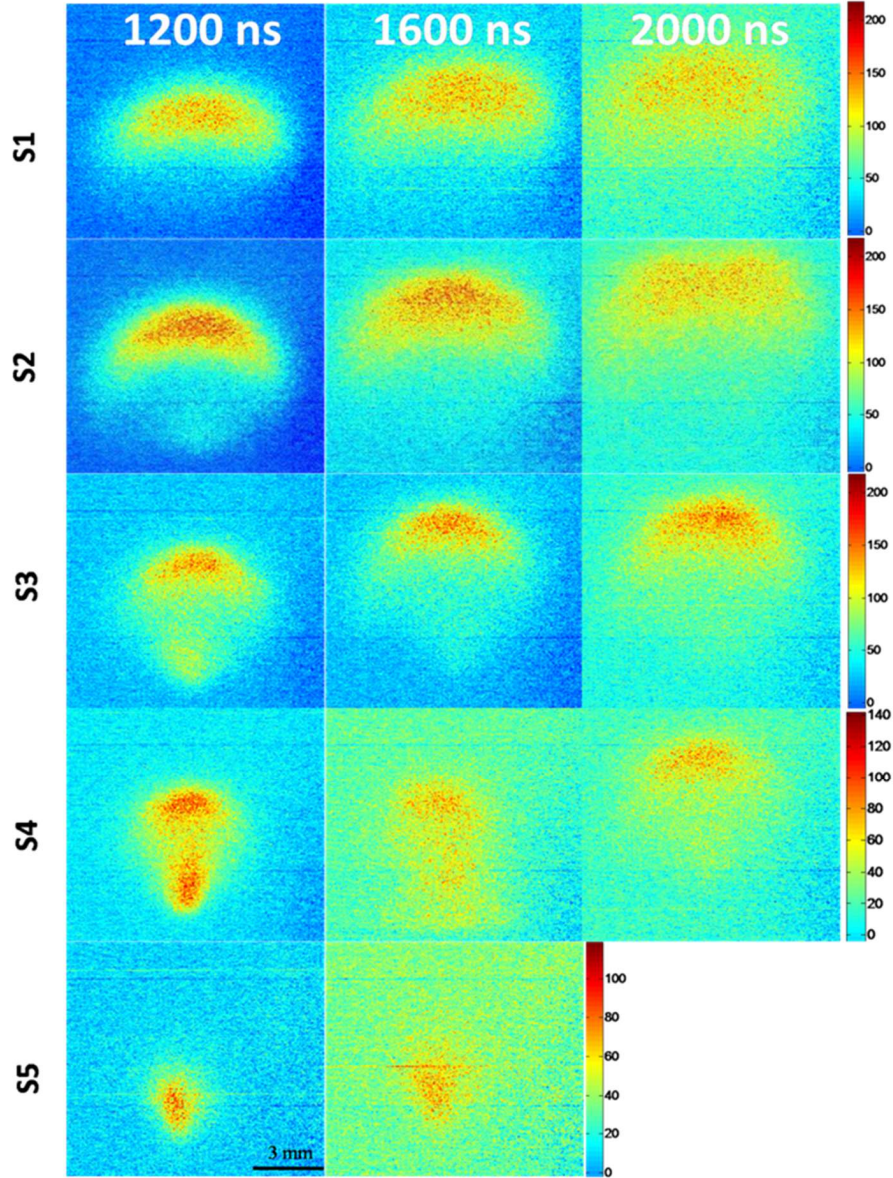

**FIG. 6.** Optical images of LPP plumes produced with different spot sizes from S1-S5 and recorded with the ICCD at different times during the late phase of plasma plume expansion *i.e.,*>1000 ns after the onset of plasma formation. Each image corresponding to a given spot size is obtained from fresh surface.

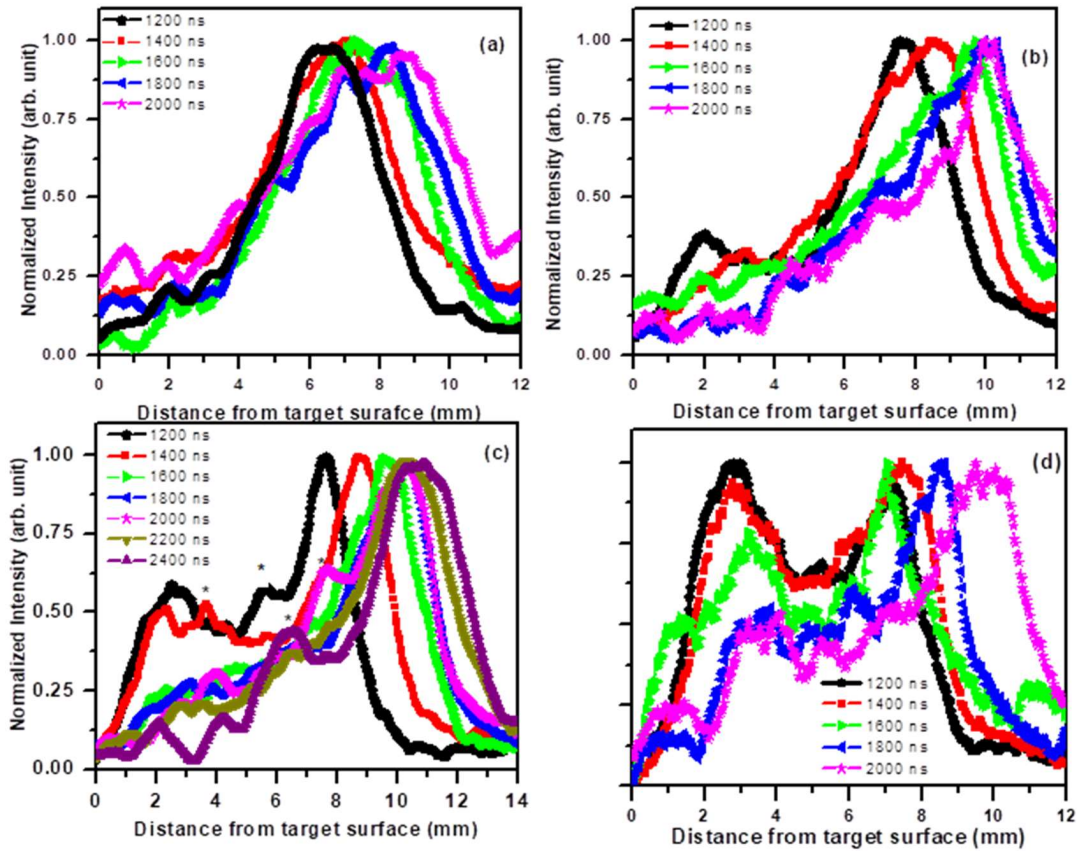

**FIG. 7.** Normalized optical intensity counts obtained from the ICCD images (Fig. 6) along the direction of plume expansion at various delays (>1000 ns) after the onset of plasma formation for S1-S5 focal spot sizes. For better results, the intensity counts from 25 pixels left to 25 pixels right on each side of the central axis of the plume are horizontally binned (inset C of Fig. 3) and the resulting intensity profiles are normalized with respect to its maximum intensity.

## [D]Size, Shape and Geometries of Plumes

For LPP in air, the position of the plume front and hence the plume length follows the shock wave model.[35,36] It is therefore reasonable to interpret plume propagation with the classic blast wave model using the expression $L(t) = (2\alpha E_0/\rho_\infty)^{1/(n+2)} t^{2/(n+2)}$; where $L$ is plasma plume length, $E_0$ is the blast wave energy, n=3, 2 or 1 for ideal spherical, cylindrical or plane wave propagation respectively, $\rho_\infty$ is the undisturbed air density and $\alpha$ is a constant.[34,37] The length and width of LPP plumes at different time delays are deduced from ICCD images utilizing the same $1/e^2$ criterion as employed in section in III.[44]

Scattered L-t data of the plumes corresponding to different spot sizes are curve fitted using the shock wave model $L(t) = at^s$; where the constant $a = (2\alpha E_0/\rho_\infty)^{1/(n+2)}$ and s= 2/(n+2), and are presented with solid lines in Fig. 8(a). The fitting parameter $s$ has a smaller value corresponding to smaller spot size showing its spherical expansion. The values of $s$ for spot sizes S1-S5 are 0.64±0.04, 0.67±0.05, 0.73±0.04, 0.72±0.06, and 0.63±0.07, respectively. Thus expansion becomes less spherical as the laser spot size increases from S1-S4. These results demonstrate that the properties of shock front propagation are strongly dependent on the laser spot size and hence the total mass ejected. At a given time, i.e. 2000 ns, the plume length is a maximum for S2, almost equal for S1 and S3, and a minimum for S5. The peak velocity $V_{Fit}$ for the front of the plasma plumes, obtained from the time derivative of the L-t curves (Fig. 8(a)), are 13.5, 24.8, 11.5, 10.0 and 6.1×10^3 m/s for the spot sizes S1-S5.

Temporal variation in the plume front position and plume length is widely exploited experimentally and theoretically,[12-21, 32-34] but there are very few reports on the lateral expansion and plume width (W), which should be diagnosed to study the temporal variation in the case of lateral confinement and its dependence on laser spot size. Scattered W-t data of the plasma plume are fitted utilizing the classical drag force model given by the expression $W = W_f[1 - \exp(-\beta t)]$; where $W_f$ is twice the stopping distance or maximum width of the plume in the lateral direction and $\beta$ is the stopping coefficient. From the above relationship, optimum fitting parameters for the focal condition S2 yielded values of $W_f$ = 9.6±0.1 mm and $\beta$ = 3.9×10^{-3}/s. It was also noted that for early times (<300 ns) lateral expansion is almost independent of laser spot size, increases abruptly and then attains saturation (Fig. 8 (b)). The actual time indices where abrupt increase and saturation are observed were strongly dependent upon the spot size and vary monotonically with it, except for the case of S5. The fitted values of $W_f$ and $\beta$ decrease with an increase in spot size from S2 to S5 and exhibit values of $W_f$= 9.6±0.1, 7.7±0.2, 6.5±0.6 and 3.3±0.3 mm with stopping coefficients 3.9, 2.6, 1.9 and 1.5 ×10$^{-3}$/s respectively, while corresponding values are 9.1±0.2 mm and $3.8 \times 10^{-3}$/s for S1. The above best fit parameters for $s$, $V_{Fit}$, $Wf$ and $\beta$ are summarized in Table II.

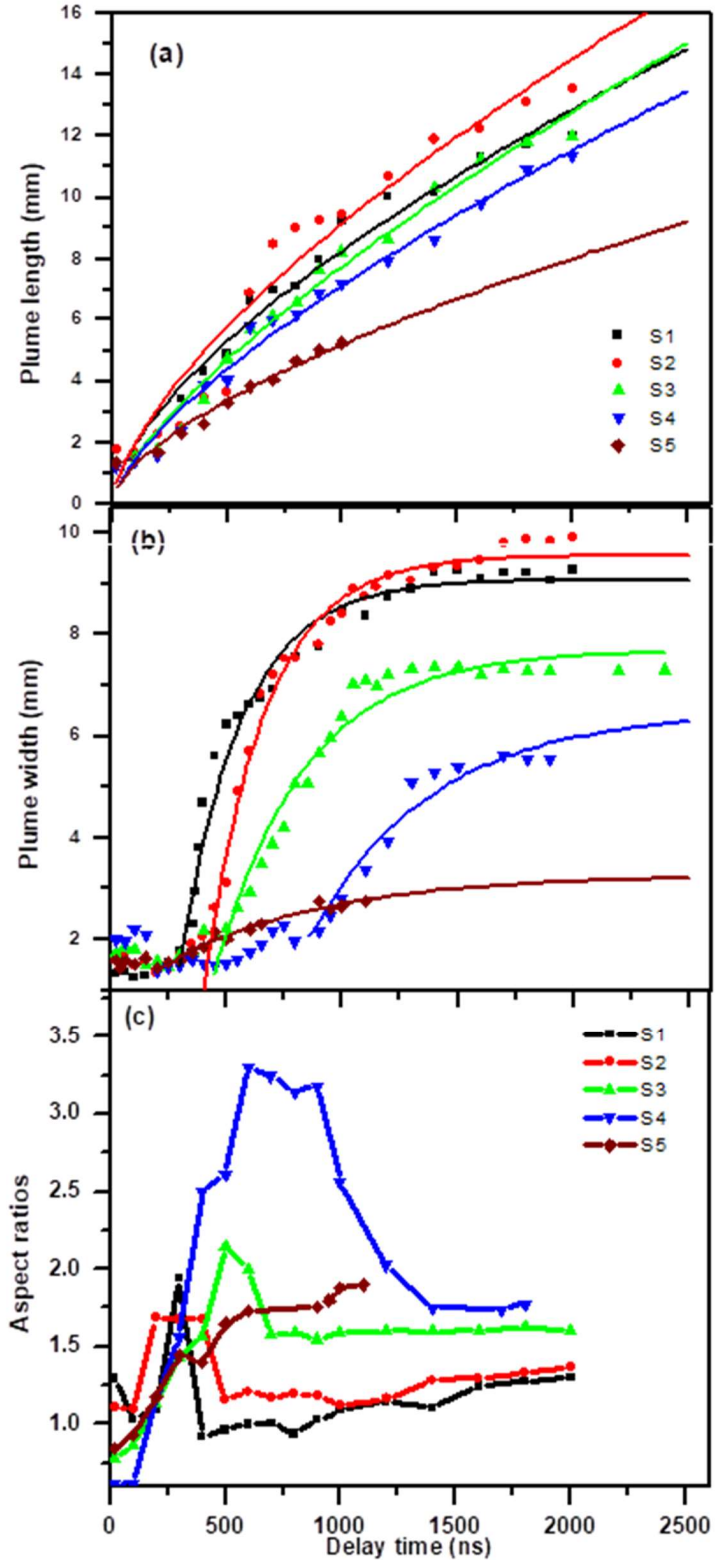

**FIG. 8.** Temporal variation in LPP plume length (a) with a function fitted using the shock model i.e., R ∝ $t^n$ , (b) Plume width and function fit using the drag model $R = R_0(1 - \exp(-\beta t))$, and (c) aspect ratio.

Aspect ratio of plasma plume is an important driver of rate of material deposition in PLD, directional laser plasma-based electron and ion sources. Aspect ratio of plasma plume increase as time progresses attains a maximum value and decreases at longer delays. Time to attain maximum aspect ratio, peak value and time width of the peak all depend on the laser spot size (fig. 8(c). With the increase of laser spot sizes from S1-S4, maximum achievable aspect ratio, time required to achieve maximum value and time range all get increased. However, for spot S5, peak aspect ratio could not achieve due to a lower plasma lifetime (1600 ns).

**Table 1I: Laser spot configuration S1-S5 plume fitting parameters: 's' parameter, $V_{Fit}$, (velocity fit), $\beta$ and $Wf$ best fit parameters.**

| Laser Spot Configuration | s | $V_{Fit} \times 10^3$ (m/s) | $\beta$ (s) | $W_f$ (mm) |
|---|---|---|---|---|
| S1 | 0.64 ±0.04 | 13.5 | 3.8 | 9.1±0.2 |
| S2 | 0.67 ±0.05 | 24.8 | 3.9 | 9.6±0.1 |
| S3 | 0.73 ±0.04 | 11.5 | 2.6 | 7.7±0.2 |
| S4 | 0.72 ±0.06 | 10.0 | 1.9 | 6.5±0.6 |
| S5 | 0.63 ±0.06 | 6.1 | 1.5 | 3.3±0.3 |

## [E] Ion Dynamics using a Langmuir Probe

Langmuir probes are widely used to determine the shape, and charge density in the flowing plasma plume. It is reported that space charge mechanisms can perturb the recorded ion signal while electrons play a vital role in space charge deformation mechanisms.[38] A Langmuir probe, positioned 2 cm from the target surface recorded ion signals from a YBCO plasma produced with different spot sizes (Fig. 9(a)). The ion TOF profiles exhibits a sharp prompt peak followed by a broader slower one. The first prompt peak originates from the photoelectric effect, where photoelectrons are generated on the probe tip from scattered laser light and initial X-ray and VUV emission during the laser pulse. It is noteworthy that narrower current traces are produced with smaller spot sizes, which agree with previous reports.[32,39] Broadening of the time-of-flight signal corresponding to larger spot sizes is a consequence of slower ions in the trailing part of the signal.

Increases in focal spot size from S1 to S2 *i.e.,* decreases in laser irradiance, increases the detected peak ion signal, while it shows an inverse relationship for variation in spot sizes from S2-S5. Time integration of the current TOF graphs (Fig. 9(a)) was used to estimate the total charge yield (TCY). It is also worth mentioning that with the increase in spot size from S1-S3, the TCY parameter increases, which indicates that spot size S3 is the most suitable for obtaining high TCY values (inset Fig. 9(a)).

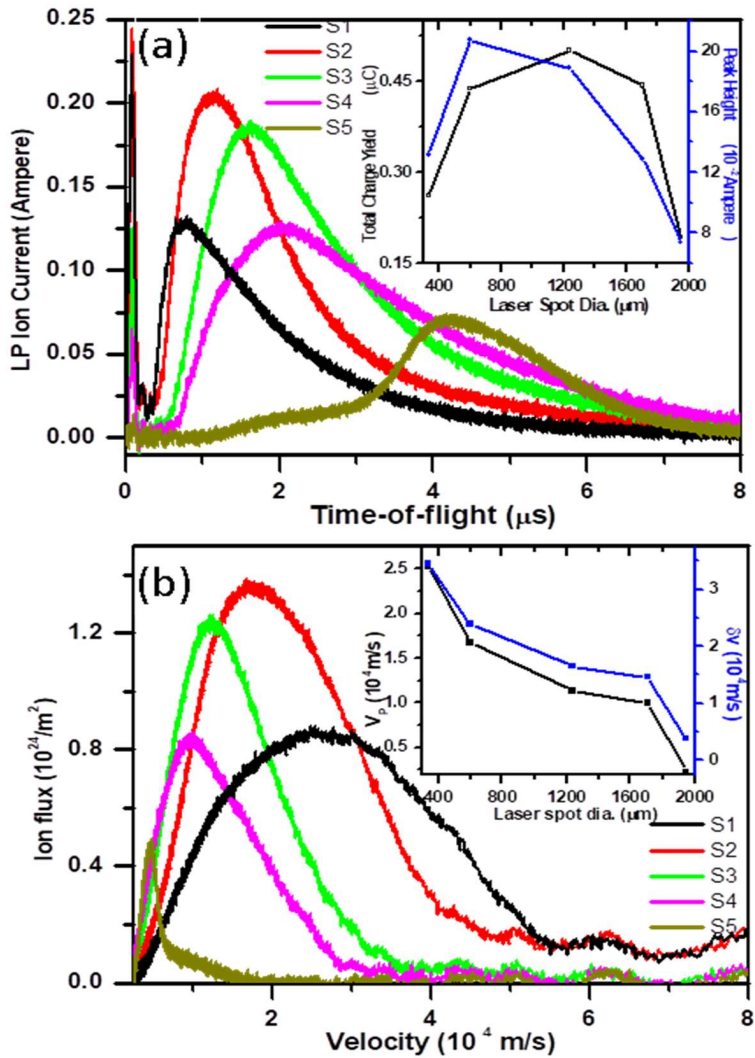

**FIG. 9.** (a) Langmuir Probe (LP) ion signals from laser produced YBCO plasmas with the probe plane situated 2 cm from target surface and facing the oncoming plume for spot sizes S1 to S5 (inset: variation of total charge yield and peak height with laser spot diameter). (b) Corresponding velocity distribution *versus* ion flux curves (inset: Variation in peak position ($V_P$) and peak width ($\delta V$) with laser spot diameter).

The velocity-ion flux graphs extracted from the LP signals in Fig. 9(a), are illustrated in Fig. 9(b). The peak expansion velocities ($v_p$) of the positive ions are estimated at 25.3, 16.8, 12.4, $9.89 \times 10^3$ and $4.78 \times 10^3$ m/s for spot sizes S1-S5 respectively, which are in good agreement with the corresponding fitted expansion velocities ($V_{Fit}$) of the plume front obtained using time derivative of the L-t profiles in Fig. 8(a). This result demonstrates that for all spot sizes diameters ranging from 0.336-1.95 mm, laser produced YBCO plasmas are predominantly populated by singly charged ions and neutral species. Laser produced tin plasmas containing more highly charged species such as $Sn^{8+}$ have been reported to exhibit higher expansion velocities for ions detected by a Faraday cup in comparison to plume front velocity recorded using ICCD images for $\Delta\lambda$=300-900 nm.[40,41] Of course, highly charged ion plasmas, which radiate predominantly in the EUV region, may yield lower plume front velocities when imaged with a visible ICCD compared to velocities from ion TOF data.[34,40,41] We observe in section III.A that the peak $V_{ESP}$ measured as well the plume width via W-t data in the previous section, decrease with increasing laser spot size. Grun et al. reported that the width ($\delta v$) of the velocity profile has a strong dependence on laser spot size, and can be controlled by a scaling parameter, $r_s/c\tau$, where $r_s$ and $\tau$ are laser spot radius and pulse duration, while $c$ is the ion sound speed.[39] Increases in spot size from S1 to S2 result in a 1.7 fold enhancement in ion flux, while a further increase from S2-S5 results in a monotonic decrease in peak ion signal. Although values of ion flux for S3 and S4 are lower than S2, they are larger compared to S1. Increases in the spot size result in a decrease in the laser fluence $F$ and hence a decrease in crater depth and volume of ablated material. [30,31]

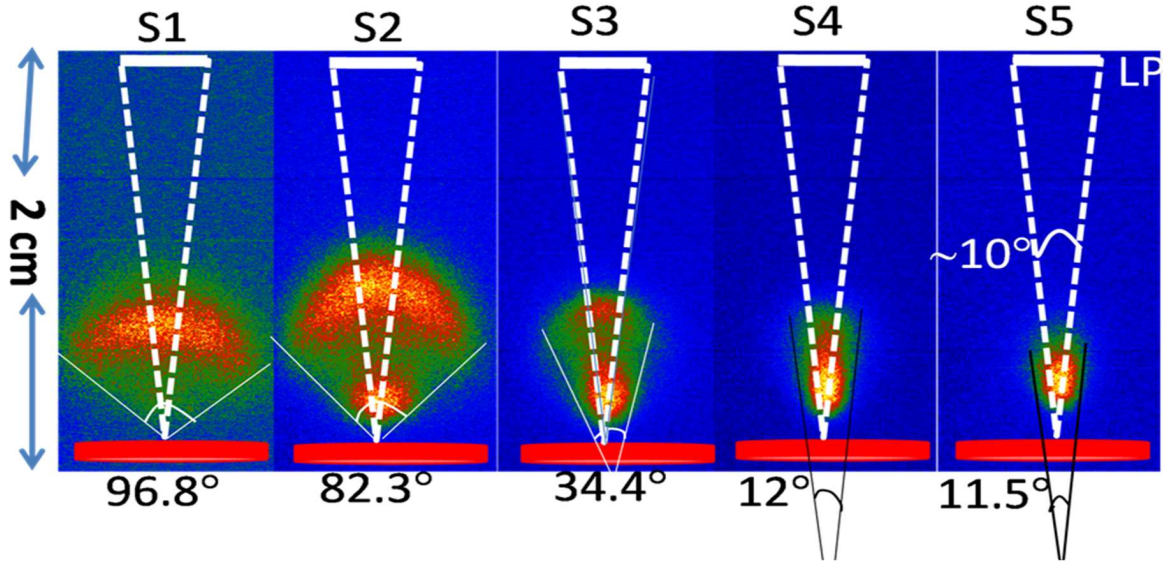

**FIG.10.** Demonstration of angular blow-off from LPP plumes at a time delay of 800 ns for different spot sizes, and their collection at the probe surface which makes a solid angle of ~10° at the center of LPP plume.

For larger spot sizes, the plasma scale length $r_s/c\tau$ is larger generating a more cylindrical plasma plume which facilities enhanced target ablation during the laser pulse duration. Plumes corresponding to smaller spot sizes expand more spherically resulting in a larger solid angle, and hence smaller fraction of ablated material can be deposited on substrate during PLD. For example at 800 ns after the onset of plasma formation, the solid angles made by the plasma plumes are 96.8°, 82.3°, 34.4°, 12°, and 11.5° for spot sizes S1-S5 (Fig. 10). The aspect ratio of the plume corresponding to spot size S4 is maximum at t > 400 ns, therefore maximum fraction of evaporated material can be collected by substrate. The surface of the Langmuir probe makes a ~10° solid angle with respect to the laser spot at the target. Ion signals registered by the LP depend upon (a) the ablated mass per pulse, (b) laser-plasma interactions(reheating) driving an increase in ion density *via* inverse Bremsstrahlung ionization of neutrals, and (c) the fraction of the ion sheath or plasma plume within the solid angle made by the LP at the target surface. Though the volume of ablated material and the reheating of this initial plasma during the laser pulse are higher for smaller spot sizes, due to the spherical expansion of the plasma, only 10 % of the plume lies in the solid angle of the LP at the target surface. In contrast, the plume and hence ion sheaths for larger spot size, e.g. for S3, exhibit a more cylindrical blow-off. Thus ~29% of the total ions can reach the probe surface, which results in almost a two-fold enhancement in TCY for S3 compared to S1. This is in accordance with the previous report of angular distributions of integrated ion yield, which demonstrated better collimation of the ion beam along the target normal for larger spot sizes.[32] Smaller values of TCYs for spot sizes S4 and S5 over that of S3 may be the consequences of smaller ion to neutral ratio and/or much lower kinetic energy of the ions. The peak ion current corresponding to S4 is slightly lower than that of S1, but the TCY in that case is almost 1.7 times higher due to its larger time width.

## IV. CONCLUSIONS

We have demonstrated the ability to monitor variation of the aspect ratio of a LPP plume, the intensity of shock waves at the plasma-air interface, the total charge yield and the width of the ion velocity distribution with varying laser spot size at the target surface. For smaller laser spot size the plume is more spherical, exhibits faster and stronger shock front formation and faster decays *i.e.* shorter emission time. Larger spot size makes LPP plumes more directional towards the substrate and therefore better suited for higher rates of deposition with minimal wastage of target material in PLD. The width of the ion velocity distribution and its peak velocity decreased with increasing spot size, demonstrating that ions corresponding to larger spot sizes are more mono-energetic.

**Acknowledgements:**

Dr. S.C. Singh is highly thankful to Irish Research Council for Science, Engineering, and Technology (now called the Irish Research Council) for providing an EMPOWER postdoctoral fellowship to work at National Centre for Plasma Science and Technology, DCU. Work supported by Science Foundation Ireland (grant nos. 12/IA/1792 and 14/TIDA/2452). This work is associated with the FP7 EU COST Action MP1208, the U.S. National Science Foundation PIRE Grant No. 1243490 and enabled by the EU FP7 Erasmus Mundus Joint Doctorate 'EXTATIC' under framework partnership agreement FPA-2012-0033.